\documentclass[12pt]{article}
\usepackage {amssymb}
\usepackage {amsmath}
\usepackage[dvips]{epsfig}

\begin{document}

\begin{center}

\Large\bf{Clustering of phosphorus atoms in silicon}
\\[2ex]

\normalsize
\end{center}

\begin{center}
\textbf{O. I. Velichko and N. A. Sobolevskaya}
\end{center}

Department of Physics, Belarusian State University of Informatics

and Radioelectronics, 6, P.~Brovka Street, Minsk, 220013 Belarus

{\it E-mail addresses: oleg\_velichko@lycos.com} (Oleg Velichko)

{\it sobolevskaya@lycos.com} (Natalia Sobolevskaya)

\begin{abstract}

To explain the effect of electron density saturation at high
phosphorus concentrations the model of negatively charged
phosphorus clusters was compared with  experimental data.  A
number of negatively charged clusters incorporating a point defect
and phosphorus atoms $(\mathrm{DP}_{1})^{-}$,
$(\mathrm{DP}_{2})^{-}$, $(\mathrm{DP}_{2})^{2-}$,
$(\mathrm{DP}_{3})^{-}$, $(\mathrm{DP}_{3})^{2-}$,
$(\mathrm{DP}_{4})^{-}$, and $(\mathrm{DP}_{4})^{2-}$ were
investigated at temperatures of 920 and 1000 $^{\circ}$C. It was
established that for clusters incorporating more than one
phosphorus atom the calculated electron density reached a maximum
value and then decreased monotonically and slowly  with an
increase in the total dopant concentration. If a cluster
incorporates 3-4 phosphorus atoms, the calculated dependencies of
carrier concentration agree well with the experimental data for
both temperatures regardless of the negative charge state of the
cluster ($-$ or $2-$). Moreover, for all  doubly negatively
charged clusters investigated the fitting parameter is independent
of the temperature. It means that formation of doubly negatively
charged clusters is most likely but it is difficult to choose
between clusters incorporating two, three or four phosphorus
atoms. From the good agreement of the calculated curves with the
experiments it follows that the model based on the formation of
negatively charged clusters can be used in simulation of high
concentration phosphorus diffusion.

\end{abstract}


\section{Introduction}

At the present time, low energy high fluence ion implantation is
widely used for producing the active regions of modern integrated
microcircuits. A combination of ion implantation having energies
of about 10 keV with rapid thermal annealing allows one to create
doped layers with a depth of about 100 nm
\cite{Whelan_00,Solmi_03}. To calculate the distributions of
electrically active dopant atoms in such small regions, models of
clustering having a high degree of accuracy should be used. The
implementation of the adequate models of clustering is very
important because  saturation of  electron density is observed
experimentally for high dopant concentrations (see, for example
\cite{Solmi_01}). Thus, only a part of dopant atoms is
electrically active, and we have difficulties in achieving high
carrier concentrations. Unfortunately, all up-to-date models
cannot adequately explain this phenomenon. Really, it is commonly
accepted that the model of Tsai {\it et al.} \cite{Tsai_80} was
the first to account for  saturation of charge carriers at an
increasing doping level \cite{Guerrero_82}.  In the model of
\cite{Tsai_80} it is supposed that arsenic clustering occurs as a
result of the reaction

\begin{equation} \label{Cluster_Reaction_Tsai}
3\mathrm{As}^{+}+e^{-}{\stackrel{\mbox{Annealing}}{\qquad
\longleftrightarrow \qquad }}
\mathrm{As}_{3}^{2+}{\stackrel{\mbox{25 $^{\circ}$C}}{\quad
\longrightarrow \quad }} \mathrm{As}_{3},
\end{equation}

\noindent where $\mathrm{As}^{ +}$  is the substitutionally
dissolved arsenic atom participating in cluster formation; $e^{-}$
is the electron.

The main feature of the model of \cite{Tsai_80} is the assumption
that clustered As atoms are electrically active (positively
charged) at annealing temperatures, while at room temperature they
are neutral. Therefore, the electron concentration at room
temperature $n_{R}$ is approximately equal to the concentration of
substitutionally dissolved arsenic atoms $C$. The assumption that
As clusters are capable of trapping the conduction electrons and
becoming neutral during cooling allows one to explain the
saturation of electron density when $C^{T}\rightarrow\infty$
\cite{Tsai_80,Guerrero_82}. Here $C^{T}=C+C^{AC}$ is the total
concentration of dopant atoms; $C^{AC}$ is the concentration of
dopant atoms incorporated into clusters. In Fig. 1, the electron
density at room temperature $n_{R}$ calculated according to the
model of Tsai {\it et al.} \cite{Tsai_80} is compared with the
experimental data obtained by Nobili {\it et al.}
\cite{Nobili_94}. The value of the fitting parameter in the mass
action law for reaction (\ref{Cluster_Reaction_Tsai}) was chosen
to satisfy the experimentally observed value of electron density
saturation $n_{max}$ which was equal to 2.11$\times$10$^{8}$
$\mu$m$^{-3}$ \cite{Solmi_01}.

\begin{figure}[ht]
\centering {
\begin{minipage}[ht]{12.0 cm}
{\includegraphics[scale=1.5]{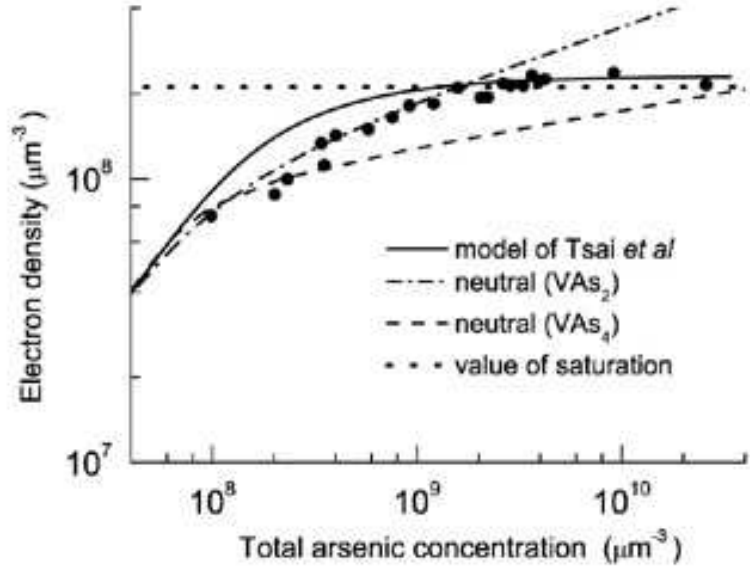}}
\end{minipage}
}

\caption{Calculated electron concentrations vs. the total arsenic
concentration for different models of clustering: solid line -
model of Tsai {\it et al.} (Ref. \cite{Tsai_80}), dash-dotted line
- neutral $\mathrm{VAs}_{2}$  clusters , and dashed line - neutral
 $\mathrm{VAs}_{4}$ clusters. The experimental data (circles) are
taken from Nobili {\it et al.} \cite{Nobili_94} for diffusion at a
temperature of 900 $^{\circ}$C. Dotted line represents the value
of saturation. \label{P_Cluster1}}
\end{figure}

As can be seen from Fig. 1, the  curve calculated disagrees with
the experimental dependence of electron density on the total
dopant concentration. In Fig. 1, the dependences $n\approx n_{R}$
calculated for the neutral clusters $\mathrm{VAs}_{2}$ and
$\mathrm{VAs}_{4}$ are also presented. Here $n=n(C^{T})$ is the
electron concentration at the temperature of cluster formation.
From the calculations presented in Fig. 1 it follows that when
neutral clusters are formed, the electron density increases
monotonically and  saturation is not observed. Thus, dependences
of electron density calculated for neutral clusters also disagree
with  experimental data.

In  \cite{Fistul'_70}, the clustering phenomenon was investigated
with the example of Ge doped by As. It was proposed that cluster
formation occurred due to the reaction

\begin{equation}\label{Reaction_Fistul'}
m\mathrm{A}^{+}  +  (m-z_{Cl})e^{ -}+
\mathrm{D}^{\times}{\longleftrightarrow} (\mathrm{AD})
^{\displaystyle z_{Cl} } ,
\end{equation}

\noindent where $\mathrm{A}^{+}$ is the substitutionally dissolved
donor atom; $\mathrm{D}^{\times}$ is the intrinsic point defect in
the neutral charge state; $m$ is the number of substitutionally
dissolved atoms participating in the reaction;
$(\mathrm{AD})^{\displaystyle z_{Cl}}$ is the cluster formed;
$z_{Cl}$ is the charge of this cluster.

It is supposed that a cluster has the same charge during annealing
and at room temperature. Based on this assumption, the dependence
between the total dopant concentration $C^{T}$ and $n_{R}$ was
obtained in \cite{Fistul'_70} from the mass action law for
reaction (\ref{Reaction_Fistul'})

\begin{equation}\label{Dependence_Fistul'}
C^{T}-n_{R}=K(T)\frac{n_{R}^{\displaystyle (m-z_{Cl})}(m \,
n_{R}-z_{Cl}C^{T})^{\displaystyle m}}{(m-z_{Cl})^{\displaystyle
m-1}} \, ,
\end{equation}

\noindent where $K(T)$ is the constant of local equilibrium for
reaction (\ref{Reaction_Fistul'}).

Two different cases of cluster formation were investigated
analytically based on  dependence (\ref{Dependence_Fistul'}): (i)
formation of  neutral clusters ($z_{Cl}$=0) and (ii) formation of
negatively charged clusters including one dopant atom and a
lattice defect. It was found that for neutral clusters the
dependence $n_{R}=n_{R}(C^{T})$ has the following form:

\begin{equation}\label{Dependence_Fistul'_neutral}
\lg n_{R}= -\frac{1}{2m}\lg (mK)+\frac{1}{2m}\lg C^{T}.
\end{equation}

It follows from expression (\ref{Dependence_Fistul'_neutral}) that
at high impurity concentrations the line with the angle
$\displaystyle \frac{1}{2m}$ in the logarithmical coordinates is
asymptotic for the dependence $n_{R}=n_{R}(C^{T})$
\cite{Fistul'_70}. This analytical result agrees with the
numerical solutions presented in Fig. 1.

In the case of a negatively charged cluster which incorporates one
impurity atom and a lattice defect Eq.(\ref{Dependence_Fistul'})
yields

\begin{equation}\label{Dependence_Fistul'_charged}
C^{T}=n_{R} \frac{1+K(T) \, n_{R}^{\displaystyle (1+
|z_{Cl}|)}}{1-\displaystyle |z_{Cl}|K(T) \, n_{R}^{\displaystyle
(1+ |z_{Cl}|)}} \, .
\end{equation}

It follows from Eq.(\ref{Dependence_Fistul'_charged}) that
saturation of electron density occurs if the dopant concentration
approaches infinity. The maximum electron concentration $n_{Rmax}$
has the following value \cite{Fistul'_70} :

\begin{equation}\label{Maximum_Fistul'_charged}
n_{Rmax}=[\, |z_{Cl}|K(T)]^{- \frac{\displaystyle 1}{\displaystyle
 1+|z_{Cl}|}} \, .
\end{equation}

Thus, the model of \cite{Fistul'_70} was the first to account for
saturation of charge carriers at an increasing doping level.

In the paper by Velichko {\it et al.} \cite{Velichko_05}, the
clustering of phosphorus atoms was investigated numerically. Due
to numerical solution, the limitations of \cite{Fistul'_70} were
removed. It was proposed that the formation of negatively charged
phosphorus clusters occurred due to the reaction

\begin{equation}\label{Reaction}
m\mathrm{P}^{+}  + m_{D} \mathrm{D}^{\displaystyle r_{D}} + ke^{
-} {\longleftrightarrow} \mathrm{(PD)}^{\displaystyle r_{Cl} } ,
\end{equation}

\noindent and that the charge of a phosphorus cluster does not
change during cooling.

\noindent Here $\mathrm{P}^{+}$ is the substitutionally dissolved
phosphorus atom; $\mathrm{D}$ is the point defect participating in
clustering; $ \mathrm{(PD)}^{\displaystyle r_{Cl} }$ is the
cluster formed; $m$, $m_{D} $, and $k$ are the numbers of impurity
atoms, defects, and electrons participating in the cluster
formation, respectively; $r_{D} $ and $r_{Cl}$ are the charge
states of the defect and cluster, respectively.

The possibility of formation of negatively charged phosphorus
clusters is indirectly confirmed in
\cite{Berding_98,Berding_Sher_98}. It follows from calculations
\cite{Berding_98,Berding_Sher_98} that singly
$(\mathrm{VAs}_{3})^{-}$ and doubly $(\mathrm{VAs}_{2})^{2-}$
negatively charged clusters can be formed at high arsenic
concentrations.

The reaction used in  \cite{Velichko_05} can be generalized to
take into account the possibility of defect generation during
clustering:

\begin{equation}\label{Reaction_Gen}
m\mathrm{P}^{+}  + m_{1} \mathrm{D}_{1}^{\displaystyle r_{1}} +
ke^{ -} {\longleftrightarrow} \mathrm{(PD)}^{\displaystyle r_{Cl}
}  + m_{2} \mathrm{D}_{2}^{\displaystyle r_{2}} ,
\end{equation}

\noindent where $\mathrm{D}_{1}^{\displaystyle r_{1}}$ and
$\mathrm{D}_{2}^{\displaystyle r_{2}}$ are the defects providing
 cluster formation and being generated during clustering,
respectively; $m_{1}$ and $m_{2}$ are the numbers of the defects
participating in reaction (\ref{Reaction}); $r_{1}$ and $r_{2}$
are the charge states of these defects, respectively. For example,
$\mathrm{D}_{2}^{\displaystyle r_{2}}$ can be the
self-interstitial generated during clustering \cite{Rousseau_98}.

The charge conservation law is valid for  chemical reaction
(\ref{Reaction_Gen}). This conservation law is of the form

\begin{equation} \label{CCL}
m + m_{1} z_{1} - k = z_{Cl}+ m_{2} z_{2}  ,
\end{equation}

\noindent where $z_{1}$ and $z_{2}$ are the charges of defects
$\mathrm{D}_{1}^{\displaystyle r_{1}}$ and
$\mathrm{D}_{2}^{\displaystyle r_{2}}$, respectively.

For reaction (\ref{Reaction_Gen}) and conservation law
(\ref{CCL}), the equations obtained in \cite{Velichko_05} to
describe clustering are modified to the following form:

\begin{equation} \label{CA}
C^{AC} = K\; \tilde {C}_{D} \;\chi ^{\displaystyle{(-z_{1}\,m_{1}+
\;k\;+z_{2}m_{2})}} \;C^{\displaystyle{m}} ,
\end{equation}

\begin{eqnarray} \label{Chi}
\chi &=& \frac{1}{2n_{i}} \left[ C + \frac{{z_{Cl}} }{{m}}C^{AC} -
C^{B}+ \right. \nonumber
\\
     &+& \left.  \sqrt {\left( C + \frac{{z_{Cl}}
}{{m}}C^{AC} - C^{B} \right)^{2} + 4n_{i}^{2}}\quad\right] {\rm ,}
\end{eqnarray}

\noindent where

\begin{equation} \label{Constant}
K = \frac{m\;K_{L} \,(h_{D1}^{
\displaystyle{r_{1}}})^{\displaystyle{m_{1}}} \,
n_{i}^{\displaystyle{(-z_{1}m_{1}+k+z_{2}m_{2})}} \,(C^{D1
\displaystyle{\times}}_{eq})^{\displaystyle{m_{1}}}} {K_{R}
\,(h_{D2}^{\displaystyle{r_{2}}})^{\displaystyle{m_{2}}} \,(C^{D2
\displaystyle{\times}}_{eq})^{\displaystyle{m_{2}}} \, },
\end{equation}

\begin{equation} \label{Rel_Defect}
\tilde {C}_{D}=\frac{(\tilde
{C}_{D1})^{\displaystyle{m_{1}}}}{(\tilde
{C}_{D2})^{\displaystyle{m_{2}}}} \, , \quad \tilde {C}_{D1}=
\frac{{C}^{D1 \displaystyle{\times}}}{{C}^{D1
\displaystyle{\times}}_{eq}} \, ,\quad \tilde {C}_{D2}=
\frac{{C}^{D2 \displaystyle{\times}}}{{C}^{D2
\displaystyle{\times}}_{eq}} \, .
\end{equation}

\noindent Here $\chi$ is the electron density normalized to the
concentration of intrinsic charge carriers in a semiconductor
during diffusion $n_{i}$; $C^{AC}$ and $C^{B}$ are the
concentration of clustered phosphorus atoms and total
concentration of acceptors, respectively. The parameter $K$ has a
constant value depending on the temperature of diffusion. This
value can be extracted from the best fit to the experimental data.
The parameters $h_{D1 }^{\displaystyle{r_{1}}}$ and $h_{D2
}^{\displaystyle{r_{2}}}$ are the constants for the local
thermodynamic equilibrium in the reactions when neutral defects
$\text{D}_{1}^{\times}$ and $\text{D}_{2}^{\times}$ are converted
into charge states $r_{1}$ and $r_{2}$, respectively; ${C}^{D1
\displaystyle{\times}}$ and ${C}^{D2 \displaystyle{\times}}$ are
the concentrations of defects $\mathrm{D}_{1}^{\times}$ and
$\mathrm{D}_{2}^{\times}$, respectively. The quantities ${C}^{D1
\displaystyle{\times}}_{eq}$ and ${C}^{D2
\displaystyle{\times}}_{eq}$ represent the equilibrium
concentrations of these defects in the neutral charge state.

In \cite{Velichko_05}, the formation of singly negatively charged
clusters incorporating one phosphorus atom or two phosphorus atoms
was investigated. It was shown that in the first case the
concentration of charge carriers reached saturation if the
concentration of substitutionally dissolved dopant atoms and
correspondingly, the total concentration of phosphorus atoms had
increased. Thus, these calculations agree with predictions of
\cite{Fistul'_70}. In the second case, the electron density
reached a maximal value and then monotonically decreased.
Unfortunately, in this paper a detailed comparison of calculated
distributions of electron density with  experimental data was not
carried out. Nor other cluster species were considered. Therefore,
the purpose of this study is to investigate systematically the
formation of different types of phosphorus clusters and compare
the calculated dependences of electron density with  experimental
data.

\section{Calculations}

The system of equations (\ref{CA}),(\ref{Chi}) obtained in
\cite{Velichko_05} and modified in this paper allows one to
consider  different types of negatively charged clusters. In the
present investigation the electron density $n_{R}\approx
n=n(C^{T})$, concentration of substitutionally dissolved
phosphorus atoms $C=C(C^{T})$, and the concentration of clustered
phosphorus atoms $C^{AC}=C^{AC}(C^{T})$ depending on the total
phosphorus concentration $C^{T}$ were calculated for the clusters
$(\mathrm{VP}_{1})^{-}$, $(\mathrm{VP}_{2})^{-}$,
$(\mathrm{VP}_{2})^{2-}$, $(\mathrm{VP}_{3})^{-}$,
$(\mathrm{VP}_{3})^{2-}$, $(\mathrm{VP}_{4})^{-}$, and
$(\mathrm{VP}_{4})^{2-}$. To calculate these functions, a
numerical solution of the system of nonlinear equations
(\ref{CA}), (\ref{Chi}) was obtained by Newton's method.

For comparison with the experimental data of \cite{Masetti_77} and
\cite{Solmi_98}, the processing temperature was chosen to be 920
$^{\circ}$C ($n_{i}$ = 5.684$\times 10^{ 6}$ $\mu $m $^{ -3}$) and
1000 $^{\circ}$C ($n_{i}$ = 8.846$\times 10^{ 6}$ $\mu $m $^{
-3}$), respectively. In the experiments carried out in
\cite{Masetti_77}, the gas composition containing 0.27 \%
$\mathrm{POCl}_{3}$ was used as a source. The diffusion was
carried out in a low dislocation density, (111) oriented, boron
doped, Czochralsky pulled silicon with a resistivity of 1 $\Omega
\;cm$. Duration of diffusion was 7 minutes at a temperature of 920
$^{\circ}$C. The distribution of carriers concentration was
extracted using the incremental sheet resistance and Hall
measurements, performed after thinning of the specimens by anodic
oxidation and oxide stripping. Neutron activation analysis was
used for measuring the total concentration of impurity atoms. In
\cite{Solmi_98}, silicon wafers, (100) oriented, p-type of 10
$\Omega$ cm resistivity, were heavily implanted with phosphorus at
a fluence of 1.0$\times 10^{17}$ cm$^{-2}$ and an energy of 100
keV and then with a fluence of 5.0$\times 10^{16}$ cm$^{-2}$ at 50
keV. The samples were furnace annealed at a constant temperature
of 1000 $^{\circ}$C for 15 min. Secondary neutral mass
spectroscopy (SNMS) was employed to measure the dopant
concentration profiles of the samples after annealing. Carrier
concentration and mobility profiles were determined by the
accurate incremental sheet resistance and Hall measurements,
performed after thinning of the specimens by anodic oxidation and
oxide stripping.

In Fig. 2, the electron density calculated for the formation of
$(\mathrm{DP}_{1})^{-}$ clusters is presented.  As can be seen
from Fig. 2, the saturation of electron density occurs at
extremely high values of the total impurity concentration $C^{T}$
and calculated dependence $n_{R} \approx n=n(C^{T})$ disagrees
with the experimental data. The disagreement is much less but also
takes place for the formation of the $(\mathrm{DP}_{2})^{-}$
clusters.

\begin{figure}[ht]
 \centering {
\begin{minipage}[ht]{12.0 cm}
{\includegraphics[ scale=1.5]{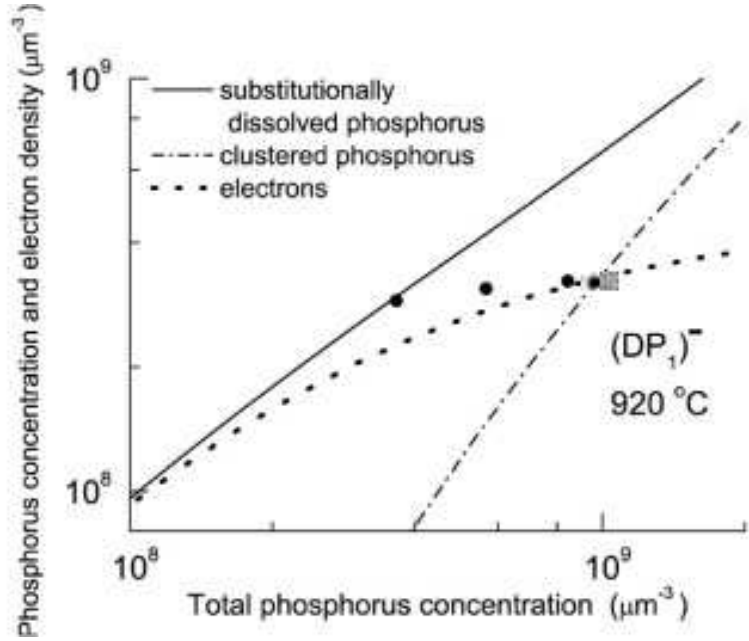}}
\end{minipage}
}

\caption{Calculated concentrations of substitutionally dissolved
phosphorus atoms (solid line), clustered phosphorus atoms
(dash-dotted line), and electron density (dotted line) vs. the
total dopant concentration when singly negatively charged clusters
incorporating a point defect and one phosphorus atom are formed.
The experimental data (circles) are taken from Masetti {\it et
al.}  \cite{Masetti_77} for  diffusion at a temperature of 920
$^{\circ}$C. \label{P_Cluster2}}
\end{figure}

\begin{figure}[!ht]
 \centering {
 \begin{minipage}[h]{8.0 cm}
{\includegraphics[ scale=1.0]{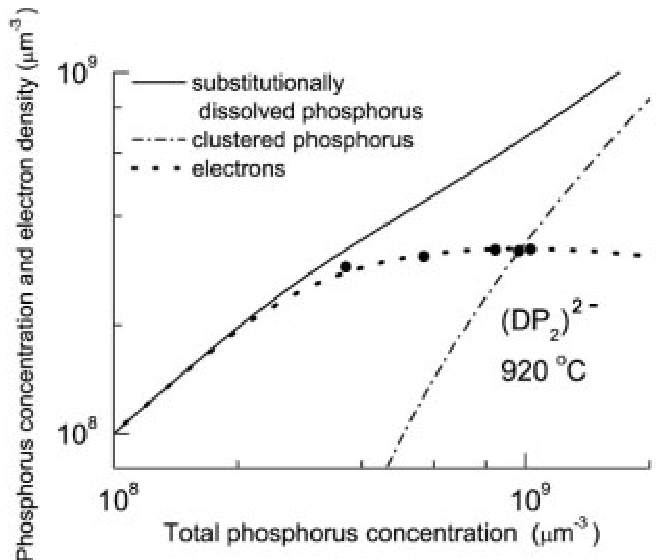}} \centerline{(a)}
\end{minipage}

\begin{minipage}[!ht]{8.0 cm}
{\includegraphics[ scale=1.0]{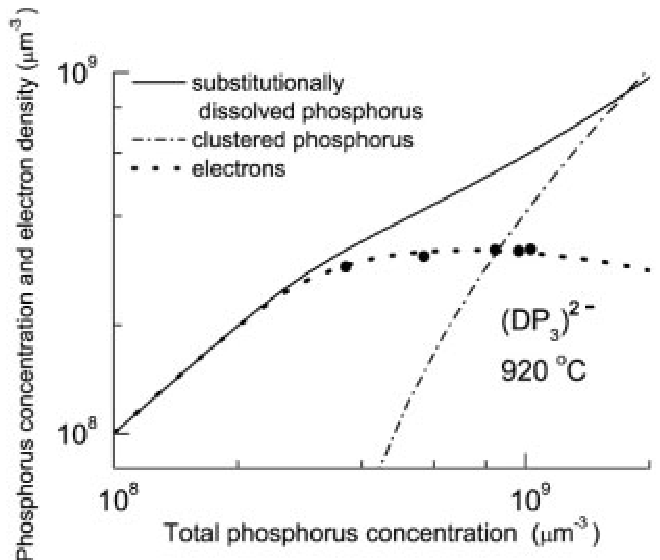}} \centerline{(b)}
\end{minipage}
}

\caption{Calculated concentrations of substitutionally dissolved
phosphorus atoms (solid line), clustered phosphorus atoms
(dash-dotted line), and electron density (dotted line) vs. the
total dopant concentration for the formation of doubly negatively
charged clusters incorporating two (a) and three (b) phosphorus
atoms. The experimental data are taken from Masetti {\itshape et
al.}  \cite{Masetti_77} for diffusion at 920 $^{\circ}$C.}
\label{P_Cluster3}
\end{figure}

The electron densities calculated for the clusters
$(\mathrm{DP}_{2})^{2-}$, $(\mathrm{DP}_{3})^{2-}$, \,
$(\mathrm{DP}_{3})^{-}$, and $(\mathrm{DP}_{4})^{-}$ are presented
in Figs. 3a, 3b, 4a, and 4b, respectively. As can be seen from
these figures, a good agreement is observed between the calculated
curves and  experimental data especially for the case of
$(\mathrm{DP}_{4})^{-}$ cluster formation. As follows from the
calculated curves, for the clusters incorporating more than one
phosphorus atom the electron density reaches a maximum value and
then decreases monotonically and slowly with increasing $C^{T}$.

\begin{figure}[!ht]
 \centering {
\begin{minipage}[!ht]{8.0 cm}
{\includegraphics[ scale=1.0]{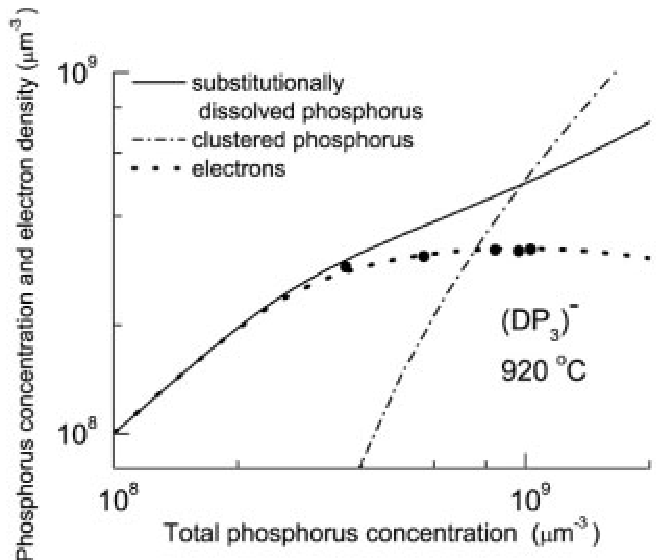}} \centerline{(a)}
\end{minipage}

\begin{minipage}[!ht]{8.0 cm}
{\includegraphics[ scale=1.0]{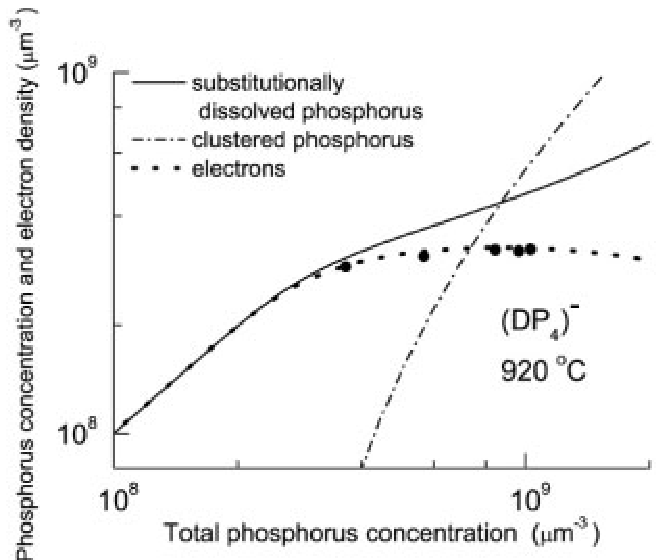}} \centerline{(b)}
\end{minipage}
}

\caption{Calculated concentrations of substitutionally dissolved
phosphorus atoms (solid line), clustered phosphorus atoms
(dash-dotted line), and electron density (dotted line) vs. the
total dopant concentration for the formation of singly negatively
charged clusters incorporating three (a) and four (b) phosphorus
atoms. The experimental data are taken from Masetti {\itshape et
al.}  \cite{Masetti_77} for diffusion at 920 $^{\circ}$C.}
\label{P_Cluster4}
\end{figure}

To calculate the electron densities presented, the generalized
system of equations (\ref{CA}),(\ref{Chi}) was transformed due to
assignment of the cluster charge state, a number of defects
participating in cluster formation (one in the cases under
consideration), and a number of phosphorus atoms incorporated into
the cluster. For example, the following equations describing the
concentration of phosphorus atoms incorporated into the
$(\mathrm{DP}_{4})^{-}$ clusters  and electron density
corresponding to the formation of these clusters can be obtained
from the system (\ref{CA}),(\ref{Chi}):

\begin{equation} \label{CA41}
C^{AC} = K\;\tilde {C}_{D} \;\chi ^{5} \;C^{\;4} ,
\end{equation}

\begin{eqnarray} \label{Chi41}
\chi &=& \frac{1}{2n_{i}} \left[ C - \frac{1}{4} K\;\tilde {C}_{D}
\;\chi ^{5} \;C^{\;4} - C^{B}+ \right. \nonumber
\\
     &+& \left.  \sqrt {\left( C - \frac{1}
{4} K\;\tilde {C}_{D} \;\chi ^{5} \;C^{\;4} - C^{B} \right)^{2} +
4n_{i}^{2}}\quad\right] {\rm .}
\end{eqnarray}

It follows from the calculated dependence $n=n(C^{T})$ that with
an increase in the total phosphorus concentration $C^{T}$ the
concentration of charge carriers $n$ reaches a maximum value
$n_{max}$ =3.25$\times 10^{8}$ $\mu $m$^{-3}$ at $C_{\max
}^{T}\sim$ \, $6.90\times $10$^{8}$ $\mu $m$^{-3}$ and then
monotonically decreases. At maximum the concentration of
substitutionally dissolved phosphorus atoms $C_{\max }$ is
$3.98\times $10$^{8}$ $\mu $m$^{-3}$. The fitting parameter
$K\tilde{C_{D}}$ used in this calculation was equal to 1.9$\times
$10$^{-35}$ $\mu $m$^{9}$.

For the case of doubly negatively charged clusters
$(\mathrm{DP}_{4})^{2-}$ a disagreement is observed between the
calculated dependence $n=n(C^{T})$ and the experimental data. It
follows from the comparison of the calculated dependences
$n=n(C^{T})$ with the experimental data that the suggestion on the
formation of negatively charged clusters $(\mathrm{DP}_{2})^{2-}$,
$(\mathrm{DP}_{3})^{2-}$, $(\mathrm{DP}_{3})^{-}$, and
$(\mathrm{DP}_{4})^{-}$ provide a good fit of calculated functions
to  experimentally measured distribution of carrier concentration
(see Figs. 3, 4). Therefore, it is difficult to extract the exact
charge state and the number of the atoms incorporated into the
cluster from the experimental data presented  but the best fit is
observed for the  $(\mathrm{DP}_{4})^{-}$ clusters.

\begin{figure}[!ht]
 \centering {
\begin{minipage}[!ht]{8.0 cm}
{\includegraphics[ scale=1.0]{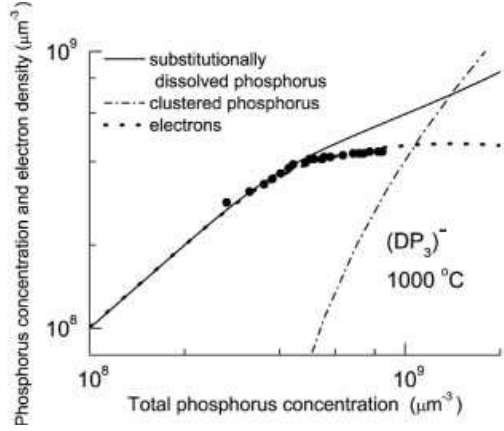}} \centerline{(a)}
\end{minipage}

\begin{minipage}[!ht]{8.0 cm}
{\includegraphics[ scale=1.0]{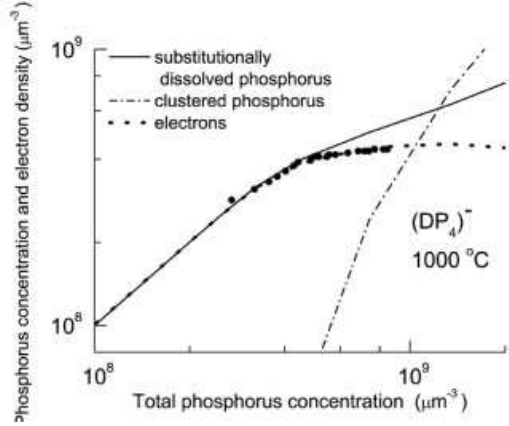}} \centerline{(b)}
\end{minipage}
}

\caption{Calculated concentrations of substitutionally dissolved
phosphorus atoms (solid line), clustered phosphorus atoms
(dash-dotted line), and electron density (dotted line) vs. the
total dopant concentration for the formation of singly negatively
charged clusters incorporating three (a) and four (b) phosphorus
atoms. The experimental data are taken from Solmi and Nobili
\cite{Solmi_98} for diffusion at a temperature of 1000
$^{\circ}$C.} \label{P_Cluster5}
\end{figure}

\begin{figure}[!ht]
 \centering {
\begin{minipage}[!ht]{8.0 cm}
{\includegraphics[ scale=1.0]{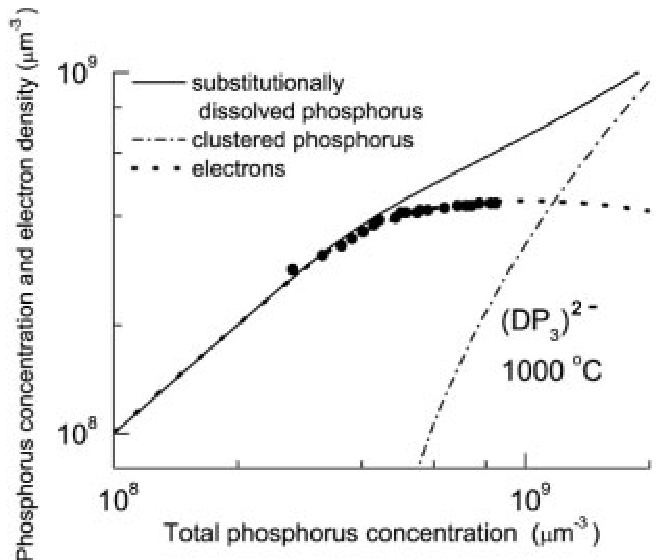}} \centerline{(a)}
\end{minipage}

\begin{minipage}[!ht]{8.0 cm}
{\includegraphics[ scale=1.0]{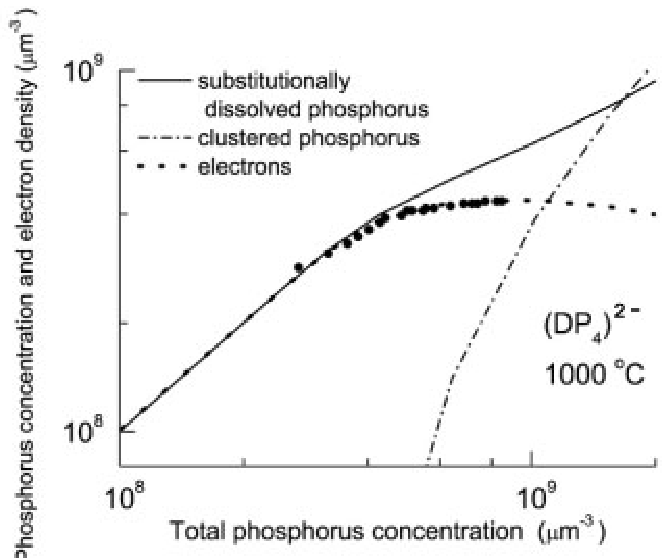}} \centerline{(b)}
\end{minipage}
}

\caption{Calculated concentrations of substitutionally dissolved
phosphorus atoms (solid line), clustered phosphorus atoms
(dash-dotted line), and electron density (dotted line) vs. the
total dopant concentration for the formation of doubly negatively
charged clusters incorporating three (a) and four (b) phosphorus
atoms. The experimental data are taken from Solmi and Nobili
\cite{Solmi_98} for diffusion at a temperature of 1000
$^{\circ}$C.} \label{P_Cluster6}
\end{figure}

In  Figs. 5a, 5b, 6a, and 6b the dependences $n=n(C^{T})$
calculated for the clusters $(\mathrm{DP}_{3})^{-}$,
$(\mathrm{DP}_{4})^{-}$, $(\mathrm{DP}_{3})^{2-}$, and
$(\mathrm{DP}_{4})^{2-}$, respectively, are compared with the
experimental data of \cite{Solmi_98} obtained for a temperature of
1000 $^{\circ}$C. As can be see from the figures, all these
clusters give a good fit to the experimental data, especially the
 $(\mathrm{DP}_{3})^{2-}$ and $(\mathrm{DP}_{4})^{2-}$ clusters.
For the case of formation of the  $(\mathrm{DP}_{3})^{2-}$
clusters, the system (\ref{CA}),(\ref{Chi}) has the form

\begin{equation} \label{CA32}
C^{AC} = K\;\tilde {C}_{D} \;\chi ^{5} \;C^{\;3} ,
\end{equation}

\begin{eqnarray} \label{Chi32}
\chi &=& \frac{1}{2n_{i}} \left[ C - \frac{2}{3} K\;\tilde {C}_{D}
\;\chi ^{5} \;C^{\;3} - C^{B}+ \right. \nonumber
\\
     &+& \left.  \sqrt {\left( C - \frac{2}
{3} K\;\tilde {C}_{D} \;\chi ^{5} \;C^{\;3} - C^{B} \right)^{2} +
4n_{i}^{2}}\quad\right] {\rm }
\end{eqnarray}

\noindent and the value of the fitting parameter $K\tilde{C_{D}}$
was chosen to be equal to 3.7$\times $10$^{-27}$ $\mu $m$^{6}$.

For the case of $(\mathrm{DP}_{4})^{2-}$ formation the system
(\ref{CA}),(\ref{Chi}) has the form

\begin{equation} \label{CA42}
C^{AC} = K\;\tilde {C}_{D} \;\chi ^{6} \;C^{\;4} ,
\end{equation}

\begin{eqnarray} \label{Chi42}
\chi &=& \frac{1}{2n_{i}} \left[ C - \frac{1}{2} K\;\tilde {C}_{D}
\;\chi ^{6} \;C^{\;4} - C^{B}+ \right. \nonumber
\\
     &+& \left.  \sqrt {\left( C - \frac{1}
{2} K\;\tilde {C}_{D} \;\chi ^{6} \;C^{\;4} - C^{B} \right)^{2} +
4n_{i}^{2}}\quad\right] {\rm }
\end{eqnarray}

\noindent and the fitting parameter $K\tilde{C_{D}}$ was chosen to
be equal to 1.65$\times $10$^{-37}$ $\mu $m$^{9}$.

As for $(\mathrm{DP}_{2})^{2-}$ cluster formation, there is a
little disagreement between the calculated electron density and
experimental data. It is interesting to note that for the case of
formation of doubly negatively charged clusters incorporating more
than one phosphorus atom the values of parameters $K\tilde{C_{D}}$
providing a best fit to  experimental data are equal for both
temperatures 920 and 1000 $^{\circ}$C. At the same time, the
values of $K\tilde{C_{D}}$ calculated for singly negatively
charged clusters are not equal for different temperatures. For
example, if a formation of $(\mathrm{DP}_{4})^{-}$ is assumed,
then the values of $K\tilde{C_{D}}$ are equal to 1.9$\times
$10$^{-35}$ and 1.25$\times $10$^{-35}$ $\mu $m$^{9}$ for
temperatures 920 and 1000 $^{\circ}$C, respectively. It follows
from the extracted values of $K\tilde{C_{D}}$ that the formation
of doubly negatively charged clusters is most likely but it is
difficult to choose between the clusters incorporating two, three
or four phosphorus atoms.

\newpage

\section{Conclusions}

Analysis of the models dealing with the dopant atom clustering in
silicon was carried out. It is shown that the  models of
clustering developed do not explain the available experimental
data, especially the effect of electron density saturation at high
dopant concentrations. However, the phenomenon of saturation can
be explained within the assumption of negatively charged
phosphorus complex formation. To investigate this model, the
electron density and concentration of dopant atoms incorporated
into clusters are calculated for temperatures of 920 and 1000
$^{\circ}$C as the functions of the total phosphorus
concentration. A number of negatively charged clusters
incorporating a point defect and phosphorus atoms,
$(\mathrm{DP}_{1})^{-}$, $(\mathrm{DP}_{2})^{-}$,
$(\mathrm{DP}_{2})^{2-}$, $(\mathrm{DP}_{3})^{-}$,
$(\mathrm{DP}_{3})^{2-}$, $(\mathrm{DP}_{4})^{-}$, and
$(\mathrm{DP}_{4})^{2-}$, are investigated. It is shown that in
the case of $(\mathrm{DP}_{1})^{-}$ formation the concentration of
charge carriers reaches saturation if the total phosphorus
concentration increases. However, the saturation occurs only for
very high values of the total dopant concentration that disagrees
with  experimental data. In the cases of clusters incorporating
more than one phosphorus atom, the electron density reaches a
maximum value and then  decreases monotonically and slowly with an
increase in the total dopant concentration. If a cluster
incorporates 3-4 phosphorus atoms, the calculated dependences of
carrier concentration agree well with the experimental data for
both temperatures regardless of the value of the negative charge
state of the cluster ($-$ or $2-$). On the other hand, for all
 doubly negatively charged clusters investigated the values of the
fitting parameter are independent of the temperature. It means
that formation of doubly negatively charged clusters is most
likely but it is difficult to choose between clusters
incorporating two, three or four phosphorus atoms. From a good
agreement of the calculated curves with  experimental data it
follows that the model based on the formation of negatively
charged clusters can be used in simulation of high concentration
phosphorus diffusion.

\end{document}